# Quasi-solid-state electrolyte for ultra-high safety and cycle stability battery

Yuewang Yang, Sijing Liu, Baoling Huang

Department of Mechanical and Aerospace Engineering, The Hong Kong University of Science and Technology, Clear Water Bay, Kowloon, Hong Kong SAR, China.

Abstract:

All-solid-state lithium batteries (ASSLB) have been regarded as the most promising candidate to achieve the next generation energy storage with high energy and high safety. However, some bottlenecks, including high interfacial resistance, bad electrochemical stability, and low conductivity, have hindered its further development. Here, we developed a $Pyr_{13}FSI/LiFSI$-based gel electrolyte and used it in the LFP/LTO full battery system to achieve a lithium-ion battery with high safety and cycle stability. The presence of ionic liquid in the electrolyte reduces the crystallinity of PVDF-HFP polymer matrix, increases the ion conductivity of the electrolyte, and greatly improves the electrode-electrolyte interface contact. These advantages enable the battery to work at room temperature and reach a specific capacity of 123mAh/g at the current of 1C. The slightly change in interfacial resistances between the gel electrolyte and electrodes with the increase of the cycle numbers is confirmed through electrochemical impedance spectroscopy. The high electrochemical stability of the electrolyte in the LFP/LTO system makes the battery exhibit good cycle stability, and the battery maintains 80% of its initial capacity after 2000 cycles at the current of 1C. In addition, benefitting from the excellent properties of ionic liquids, such as non-flammability, negligible vapour pressure, and high conductivity, the obtained gel electrolyte based LFP/LTO pouch battery exhibits high safety and cycle stability.

# Introduction

Lithium-ion batteries (LIBs) with the advantages of high energy and cycle stability have occupied an absolute dominant position in the field of energy storage applications, such as portable devices, electric vehicles, since its commercialization by Sony Corporation in 1991[1]. Current commercial LIBs use organic liquid electrolytes (lithium salts dissolved in carbonate solvents), which have significant advantages of high conductivity and excellent wettability on electrodes surfaces[2]. However, high temperature caused by short circuit, overcharging, abuse, or other failure mechanism will decompose organic electrolyte into combustible gas, including $CO$, $CH_4$, $C_2H_4$, $C_2H_6$, $C_3H_6$ and so on, which cause security issues (fire and explosion)[3]. In addition, thermally induced degradation, including heat generation and gas evolution, is serious for commercial graphite anode at high temperature due to breakdown of the solid electrolyte interphase (SEI) and reaction between the lithiated carbon and electrolyte[4–6]. Safety accidents caused by battery failures in portable devices and electric vehicles have been happening all the time, seriously threatening people's safety and hindering further expansion of large-scale implication for LIBs. Therefore, the development of a safer LIBs system is urgent.

Solid-state polymer electrolytes with superior features such as high safety, no leakage, non-flammability, good flexibility, and thermal stability[7] have received a huge of attention in achieving high-safety LIBs since the first PEO/Li based solid polymer electrolyte found in 1978 by Armand[8]. Though the advantages mentioned above, the commercialization of polymer electrolyte is still out of reach due to the low conductivity and high electrolyte-electrode interface resistance. Therefore, gel polymer electrolytes were developed using liquid electrolytes as the plasticizers to increase ion conductivity and improve interface contact[9]. Currently, PEO[10], PMMA[11], PAN[12], PVDF[13], PVDF-HFP[14] are commonly used polymer matrix for gel electrolytes. Among them, PVDF-HFP with high dielectric constant (=8.4), good mechanical properties, and good liquid uptake ability has become a popular candidate[14]. On the other hand, the commonly used plasticizers are flammable organic electrolytes, which make gel electrolytes still unsafe to a certain degree. To completely solve the potential safety hazards, while maintaining good conductivity and improved interface contact, ionic liquids (ILs)-based gel electrolytes were developed[15] due to the outstanding characteristics (negligible volatility, non-flammability, thermal and electrochemical stability)[16] of ILs.

The ILs are entirely composed of ions and maintain a liquid state at room temperature. The properties of the ILs are easy to be adjusted through changing the species of the anions and cations. Room temperature ionic liquids (ILs) have attracted extensive research as promising electrolytes for batteries and super capacitors due to their high safety and high stability. Common cations include pyridinium (Py), imidazolim (Im), pyrrolidinium (Pyr), ammonium[17]. Among them, Im based ionic liquids have the highest conductivity but are not chemically stable enough compared with others due to the double Π bond in the structure, while Pyr based ionic liquids are chemically stable and have acceptable conductivity due to the low viscosity. Common anions, including $PF_6^-$, $TFSI^-$, $FSI^-$, $BF_4^-$, and $BOB^-$, are the same as used in lithium salts and its properties are consistent with those shown in corresponding lithium salts. Commercial $LiPF_6$ is thermally instable and moisture sensitive, decomposing to produce $PF_5$ through $LiPF_6(s) \Rightarrow LiF(s) + PF_5(g)$, which in turn reacts with solvents to generate highly toxic substances[17]. In contrast, LiFSI exhibits stability towards hydrolysis and high temperature. In addition, its conductivity is higher than other salts due to its small size. The order of conductivity of different salts follows $LiFSI > LiTFSI > LiClO_4 > LiBF_4$ in EC/EMC solution[18,19].

In addition to the safety issues caused by electrolytes, the electrodes also need to be taken into consideration. Commercial graphite anode has low operating voltage (below 0 V vs Li+/Li), which results in no electrolyte that can be stabilized with the graphite anode and inevitable interfacial reaction. The high activity of the graphite anode brings about the continuous growth of solid electrolyte interphase (SEI) during the cycling process and safety problems of lithium dendrite growth[20]. The formation of SEI consumes the lithium and reduces the capacity and coulombic efficiency of the battery. What's more, continuous growth of dendrites will pierce the separator, causing a short circuit and eventually developing into a fire[21]. To overcome the mentioned problems of the anode materials, various anode materials with improved specific capacities and stability have been proposed for lithium-ion batteries, such as Si-based[22], Sn-based[23], hard carbons[24] and $Li_4Ti_5O_{12}$ (LTO)[25]. Among them, LTO has been regarded as the most advantageous competitor, because it has negligible volume change during lithiation and delithiation, excellent cycle reversibility, and none solid-electrolyte-interphase (SEI) due to the high plateaus (1.55V vs Li/Li+), which make the LTO anodes can guarantee higher safety and cycle stability than graphite anodes[25]. Though mentioned advantages, its high working plateaus renders the overall energy density low in the full battery when it is used as anodes.

Holding the goal of thoroughly realizing battery safety, we designed a full battery with PVDF-HFP/Pyr$_{13}$FSI/LiFSI based gel as electrolytes, LTO as anodes, and LFP as cathodes. The Pyr$_{13}$FSI/LiFSI solution was chosen to form gel polymer electrolyte (GPE) with PVDF-HFP for good chemical stability, thermal stability, and high ion conductivity. The obtained GPE possesses high conductivity (3.3 mS/cm) at room temperature, wide electrochemical window, and good incombustibility. And the demoed full battery exhibits acceptable rate performance and outstanding cycle stability at room temperature. Additionally, the full battery shows super security due to the high safety of electrolyte and LTO anode materials.

## Experimental

### Materials

Poly(vinylidenefluoride-co-hexafluoropropylene) (PVDF-HFP, Sigma Aldrich, Mw = $4 \times 10^5$), poly(vinylidene fluoride) (PVDF, Kelude, Mw=$1\times10^5$), lithium titanate (LTO, Kelude, 459.1448 g/mol), lithium iron phosphate (LFP, Kelude, 157.76 g/mol), lithium bis(fluorosulfonyl) imide (LiFSI, Kelude, 187 g/mol), lithium bis(trifluoromethanesulfonyl) imide (LiTFSI, Kelude, 287 g/mol), carbon black (CB, Kelude, 12g/mol), N-propyl-N-methylpyrrolidinium bis(fluorosulfonyl) imide (Pyr$_{13}$FSI, Chengjie, 308.37g/mol), N-butyl-N-methylpyrrolidinium bis(trifluoromethanesulfonyl) imide (Pyr$_{14}$TFSI, Chengjie, 422 g/mol). All chemicals were used as received.

### Synthesis and characterization of the GPE

The GPE was synthesized by a simple casting method. Firstly, PVDF-HFP was dissolved in acetone (concentration: 0.1 g/ml) followed by stirring for 3 hours to form a homogenous and transparent solution. Meanwhile, LiFSI was dissolved in Pyr$_{13}$FSI to get a liquid electrolyte with the concentration of 1 mol/L and stored in the glove box. Then, the liquid electrolyte was added into the PVDF-HFP solution with the mass ratio between the liquid electrolyte and PVDF-HFP of 3:1. After 3 hours of stirring, casting the solution to a petri-dish and heated for 2 hours at 50°C to form a transparent membrane. Finally, baked the membrane under vacuum (<0.1 MP) at 80 °C for ten hours to remove the remaining water and acetone. Then quickly transfered to the glove box for further use. The control group (traditional organic electrolyte-based gel electrolyte) was prepared by the phase-separation method. Firstly, PVDF-HFP, glycerol, and DMF were mixed at the weight ratio of 1:2:10. Then, evaporated the DMF under vacuum at 80 °C for 12 hours to get the preliminary porous membrane. The membrane was further immersed into DI water to extract glycerol to form micropores. Finally, the gel electrolyte was obtained by soaking the membrane in the electrolyte of 1 M LiPF$_6$ in EC/EMC.

Thermal stability of the GPE was studied by Thermogravimetric Analysis (TGA) in Nitrogen up to 600 with a heating rate of 10 °C/min using a Netzsch STA 409. FTIR spectra for GPE were recorded on a

Perkin Elmer Spectrum 100 spectrometer in the 650–1650 $cm^{-1}$ range using the Attenuated Total Reflectance (ATR) accessory with a resolution of 2 $cm^{-1}$.

Electrochemical measurements

The ion conductivity of the gel electrolyte was measured through electrochemical impedance spectroscopy (EIS) test using symmetric stainless steel (SS)/GPES/(SS) cell, and the frequency range from 0.1 Hz to 1MHz with an amplitude of 10 mV. The electrochemical window of the gel electrolyte was measured through linear sweep voltammetry (LSV) method using Metrohm Autolab PGSTAT302N electrochemical workstation. Stainless steel used as working electrode, lithium foil used as reference and counter electrode. Two same coin cells were assembled using same electrolytes. One scanned from open circuit voltage to 6 V, and the other scanned from open circuit voltage to -1 V. Cyclic voltammograms (CV) of the full battery were recorded using Autolab workstation at the scan range of 0.8-2.8 V with different scan rate. Electrochemical impedance spectrum (EIS) analysis of the full battery was tested in the frequency range of $0.1-10^5$Hz with a 10mV perturbation.

Battery assembly and test

Cathode or anode active materials, carbon black, and poly (vinylidene fluoride) binder were dispersed in NMP solution in a weight ratio of 92:3:5 (wt.%). The raw materials were then uniformly mixed using planetary ball milling with the speed of 300 r/min for 2 hours. The obtained slurries were coated on the carbon coated Al foil current collector and dried overnight in the vacuum oven at 393 K (100 ºC) to get the electrodes. Then the electrodes were compressed using a roll squeezer to increase the compaction density and the contact between active materials and the foil. The mass loadings of cathode and anode are around 5.5 $mg/cm^2$ and 4.8 $mg/cm^2$, respectively, and the anode is the limited electrode. The pouch cell batteries were assembled in a glove box (water content<0.1 ppm, oxygen<0.1 ppm) using GPEs, LTO anode and LFP cathode. All the batteries were tested on xinwei cell test system and cycled between 0.8-2.8 V for different currents at room temperature.

## Results and discussion

The GPEs with different mass ratio of IL solution and PVDF-HFP were synthesized by a simple casting method (Fig. 1a). Due to the good miscibility between PVDF-HFP and $Pyr_{13}FSI$, the GPE exhibits high optical transmittance (Fig. 1b), which reaches a high value of 90% in the visible wavelength region from 400 to 800 nm. And the photograph of the transparent GPE is presented in the inset of Fig. 1c and the HKUST logo underneath the GPE can be clearly observed. The ionic liquid acts as the plasticizer in the gel electrolyte and turns crystalline dominated polymer into amorphous dominated polymer, which facilitates the transport of ions. The FTIR results confirm transformation from

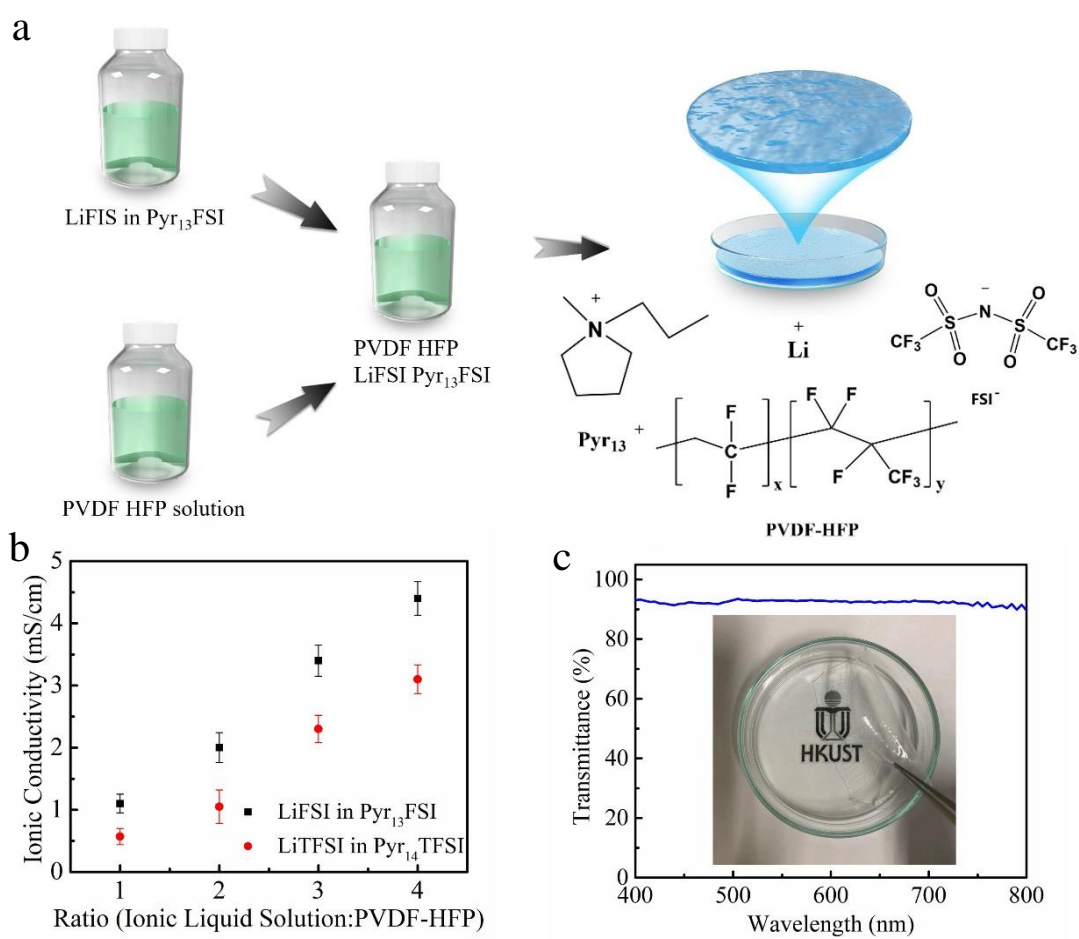

Fig. 1 (a) Schematic of the process for fabrication of GPEs. (b) Comparison of the ion conductivity of $FSI^-$-based and $TFSI^-$-based GPEs in different mass ratios. (c) Transmittance test of the GPE in the visible range.

crystalline (α phase) to amorphous (β phase) for PVDF-HFP after adding ionic liquid solution (Fig. S1 and Fig. S2). In addition, low viscosity of Pyr$_{13}$FSI and small size of LiFSI make the final GPE have high conductivity, reaching 3.3 mS/cm at the mass ratio of 3:1, as shown in Fig. 2b and Fig. S3. We make a comparison with the commonly used TFSI$^-$ anion ILs and find that the ion conductivity of FSI$^-$ based are always higher than the that of TFSI$^-$ of the same proportion.

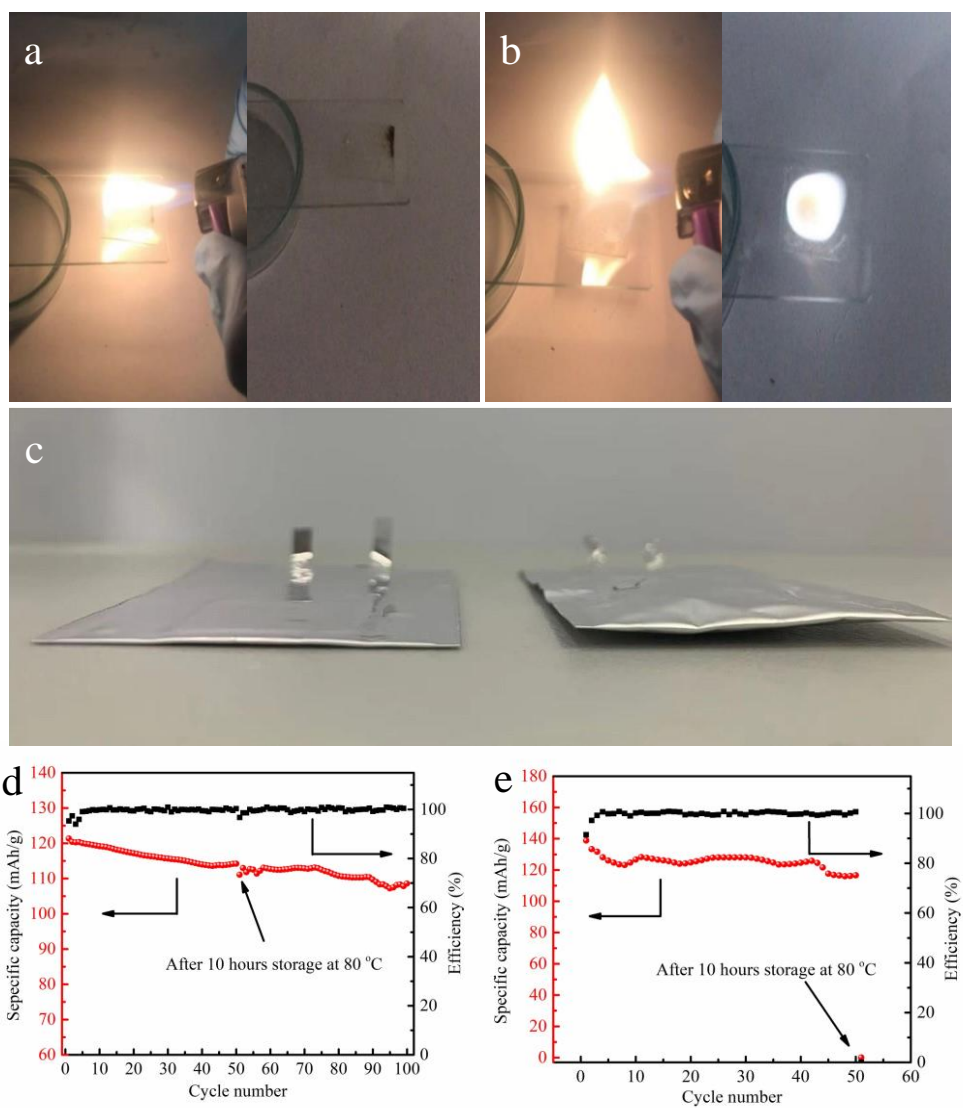

Fig. 2 (a) Flammability of the ionic liquid based GPE. (b) Traditional organic electrolyte based gel electrolyte. (c) Volatility comparison, after storing the pouch cell batteries at 80°C for 10 hours (left: GPE-based battery, right: organic electrolyte-based battery). (d) Cycle stability test for GPE-based pouch cell battery before and after 80 °C storage for 10 hours. (e) Cycle stability test for organic electrolyte-based pouch cell battery before and after 80 °C storage for 10 hours.

To characterize the high safety of GPE, the flammability tests of the electrolytes were conducted. When exposing to fire, the GPE does not catch fire at all (Fig. 2a). However, the traditional organic based gel electrolyte is still burning even after we removed the lighter (Fig. 2b). Thus, highly improving safety is achieved using GPE compared with traditional organic electrolyte-based gel electrolytes. This flammability test confirms the non-flammability of the GPE and provided a basis for the realization of a safe battery. In addition, the GPE exhibits excellent non-volatility, which further ensures high safety. After baking the pouch cell battery at a high temperature of 80 °C for 10 hours, the appearance of the GPE based battery remains the same as the initial state (Fig. 3c). However, the traditional organic electrolyte-based battery swells due to the volatilization of the organic electrolytes inside. Besides the obvious difference in appearance, there is also a significant difference in the cycle stability of the batteries. The GPE-based pouch cell battery exhibits the same specific capacity (around 120 mAh/g) and cycle stability before and after 80 °C storage (Fig. 3e). In contrast, the electrodes have been peeled off due to the volatilization of the electrolyte for the traditional organic electrolyte-based battery and the battery fails completely after 80°C storage (Fig. 3f). The specific capacity (120 mAh/g) of the GPE-based pouch cell battery is slightly smaller than the specific capacity (135 mAh/g) of the traditional organic electrolyte-based battery, because the GPE has lower conductivity and poor electrode-electrolyte interface wettability compared with traditional organic electrolytes.

To further confirm the thermal stability of the GPE, thermogravimetric analysis (TGA) was performed from 30 °C to 600 °C at a heating rate of 10 °C/min under $N_2$ (Fig. 3b). The traditional organic electrolyte is easy to volatilize and lose 80 % of its weight before 150°C. However, there is almost no weight loss

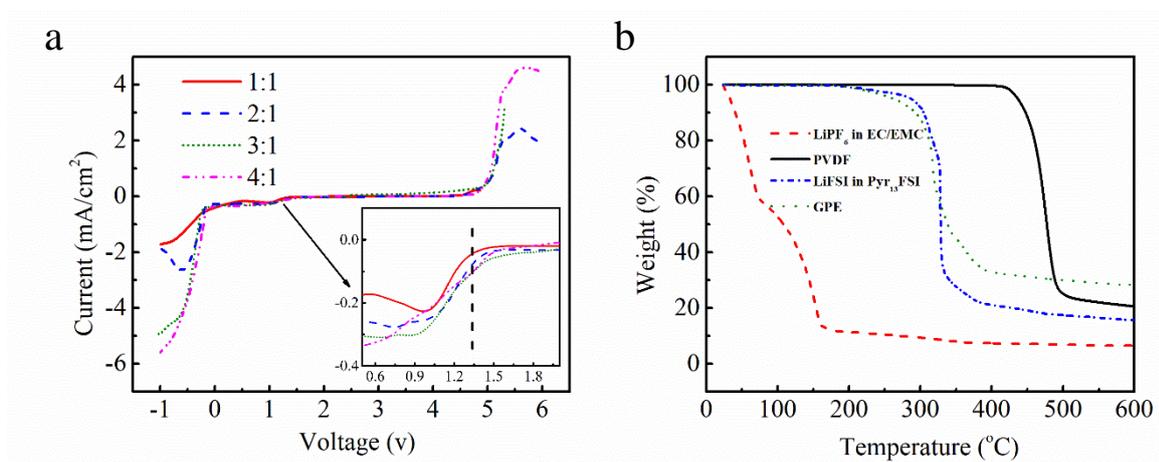

Fig. 3 (a) LSV tests of different mass ratios between ionic liquid and PVDF-HFP, the test range is from -1V to 6V. (b) TGA tests of organic electrolyte, PVDF-HFP, LiFSI in $Pyr_{13}FSI$, and GPE.

below 200 °C for ionic liquid solution, PVDF-HFP, and GPE due to the low vapour pressure of the ionic liquid and the thorough removal of the cosolvent (acetone). The TGA results prove the high thermal stability of the GPE compared with the traditional organic electrolyte. The samples containing $Pyr_{13}FSI$ begins to slowly lose weight at 200°C and lose weight sharply above 300°C. The $FSI^-$ anion decomposes at 183°C and the possible decomposition reaction is proposed by Huang and Hollenkamp[26,27].

In addition to considering the high safety of the electrolyte itself, the electrochemical stability of the electrolyte when in contact with the electrode materials is also very critical. The electrochemical stabilities towards oxidation and reduction of the GPE are investigated by LSV as shown in Fig. 4b. The GPEs with different ionic liquid contents show basically the same electrochemical stability. There is no obvious significant oxidation peak until 5 V, indicating that the obtained GPEs are electrochemically stable up to 5 V (vs. $Li/Li^+$). For the cathodic scan, a small peak can be observed at around 1.4 V vs. $Li/Li^+$, which corresponds to the cathodic limiting potential. The wide electrochemical window, 1.4-5 V vs. $Li/Li^+$ covers the working potential for LFP/LTO system (1.55-3.5 V vs $Li/Li^+$), indicating that the GPE is completely electrochemically stable to the LFP/LTO battery system. What's more, since the cathodic limiting potential of the GPE is smaller than that of the LTO, the electrochemical-based SEI is not existed in this LFP/GPE/LTO system. The absence of the electrochemical-based SEI means less lithium source consumption, high cycle stability, and high safety.

To evaluate the performance of the GPE in the battery, an LFP/LTO pouch cell battery was assembled using GPE, in which the weight ratio of ionic liquid solution and PVDF-HFP is 3 to 1. Batteries using polymer-based all-solid-state electrolytes cannot work at room temperature due to the low conductivity of the electrolytes and large electrolyte-electrode interfacial impedance, which greatly Fig. 4 (a)

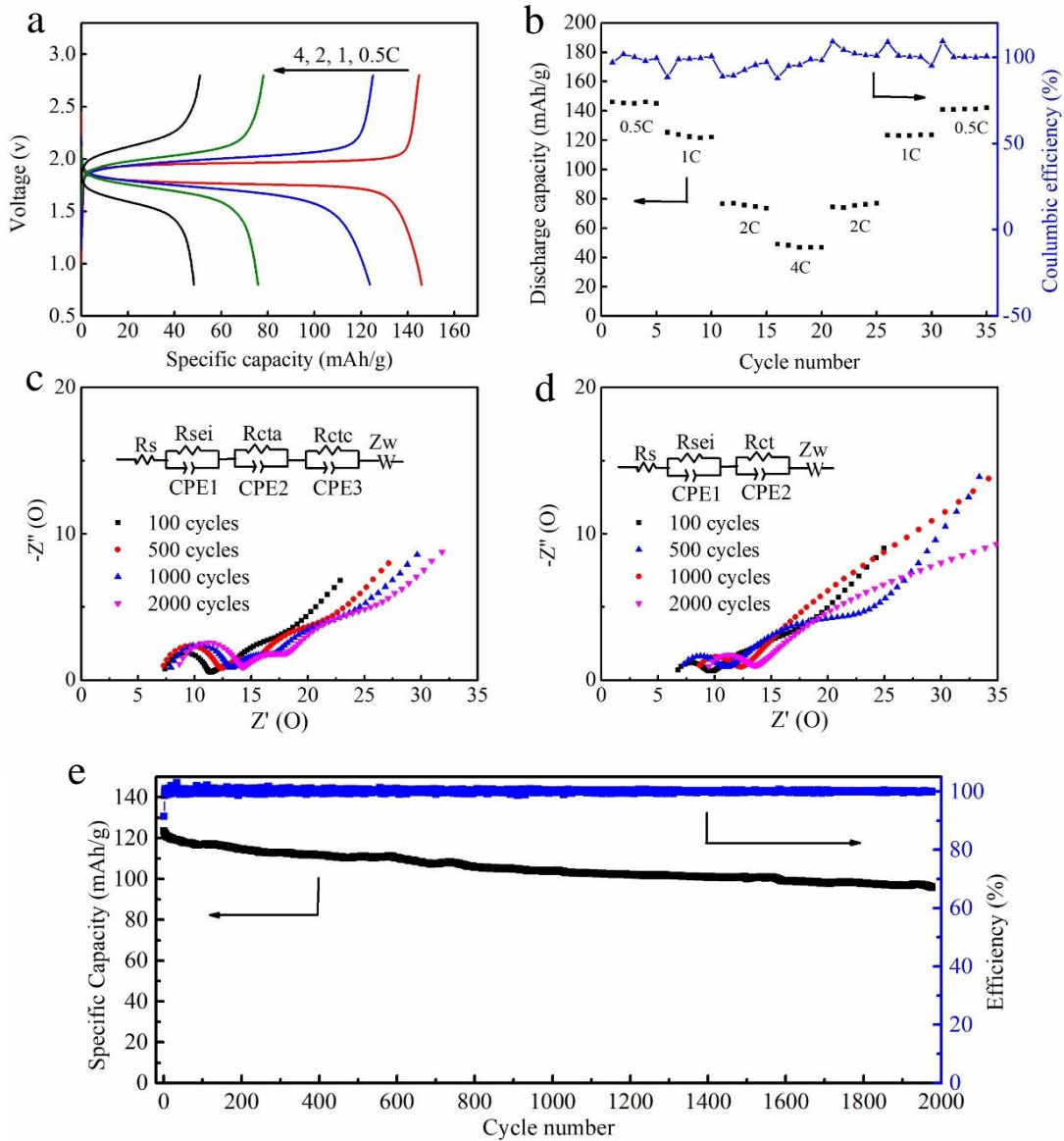

Discharge and charge profiles of the full batteries at different rates. (b) Specific capacities of the full battery at different rates. (c) EIS of GPE-based battery after different cycles. (d) EIS of traditional organic electrolyte-based battery after different cycles. (e) Cycling performance test at 1C for 2000 cycles.

limits its practicability. The ionic liquid in the GPE leads to high conductivity (3.3 mS/cm) of the electrolyte and improves the electrolyte-electrode interface contact, making the battery can work well at room temperature. Fig. 4a shows the specific capacities of the GPE based full battery are 145, 123, 76, and 49 mAh/g at the currents of 0.5C, 1C, 2C and 4C, respectively, at room temperature. The specific capacity we measured at low current (0.5C) is close to the theoretical specific capacity for the LFP/LTO full battery, which is 170 mAh/g. The specific capacity is still acceptable at the current of 1C, reaching 123 mAh/g. In addition, the specific capacity is measured again after running at high current density (4C) and it still reaches 143 mAh/g (0.2C), which means that the battery has good stability and negligible deterioration under high current. The gap between charging and discharging voltage platform obviously increases as the current increases due to the increased polarization (Fig. 4a). This polarization is also verified in the cyclic voltammetry (CV) test, as shown in Fig S5, the gap between anodic current peak and cathodic current peak increases from 0.3 to 0.5 V as the scan rate increases from 0.1 mV/s to 0.5 mV/s. At low scan rate of 0.1 mV/min, the peaks at 2.04 V and 1.70 V correspond to the insertion and extraction of lithium ions from LTO.

The long cycle stability of the pouch cell full battery at room temperature is shown in Fig. 4e. The specific capacity still reaches 98 mAh/g after 2000 cycles at 1C, which is the 80% of the initial specific capacity (123 mAh/g). The outstanding capacity retention comes from the high electrochemical stability of the GPE and the good matching between the GPE and electrode materials. In addition, the Coulombic efficiency of every cycle is close to 100%, indicating the good reversibility at charging and discharging process. The electrochemical stability of the GPE in LFP/LTO full battery have been theoretically analysed through LSV in the previous results. Here, the interfacial stability between the GPE and LTO, LFP electrodes is evaluated by the interfacial impedance analysis. Electrochemical impedance spectroscopy (EIS) spectra of the full batteries are investigated after different cycles, and the EIS test is performed at a frequency range of $10^5$ and $10^{-1}$ HZ when the battery is fully discharged and standing for half an hour to reach a stable open circuit potential. The comparison of EIS for the full batteries using GPE and traditional organic electrolyte after different cycles are shown in Fig. 4c and Fig. 4d. The insets in Fig. 4c and Fig. 4d are the equivalent circuit models used to fit the EIS data and the fitting results are given in Table S3 and Table S4. The internal resistance (Rs, high frequency intercept with real axis) is determined by the ionic resistance of the electrolyte, the intrinsic resistance of active materials, and the contact resistance between active materials and current collectors. The interfacial resistance (Rsei, high frequency semicircle) is determined by the solid electrolyte interface between LTO and electrolytes. The charge transfer resistances for anode (Rcta, medium to low frequency semicircle) and for cathode (Rctc, medium to low frequency semicircle) represent the kinetic resistance of the charge transfer at the boundary between electrodes and electrolytes. The Rs of the GPE-based full battery is slightly higher than traditional organic electrolyte-based one due to the lower conductivity of the GPE compared with liquid electrolytes. The Rsei, Rcta, and Rctc of the GPE-based full battery changes very little after 100, 500, 1000, and 2000 cycles, which proves the good stability of the solid electrolyte interface, charge transfer among the electrolyte and electrodes during the cycle (Table S3). In contrast, the Rsei and Rct of the organic electrolyte-based full battery change greatly, which means the interfaces in the battery varies greatly and the system is unstable (Table S4). Overall, the EIS tests confirm the high stability of our GPE in the LFP/LTO system, and the GPE plays a decisive role in improving cycle life and safety of the battery.

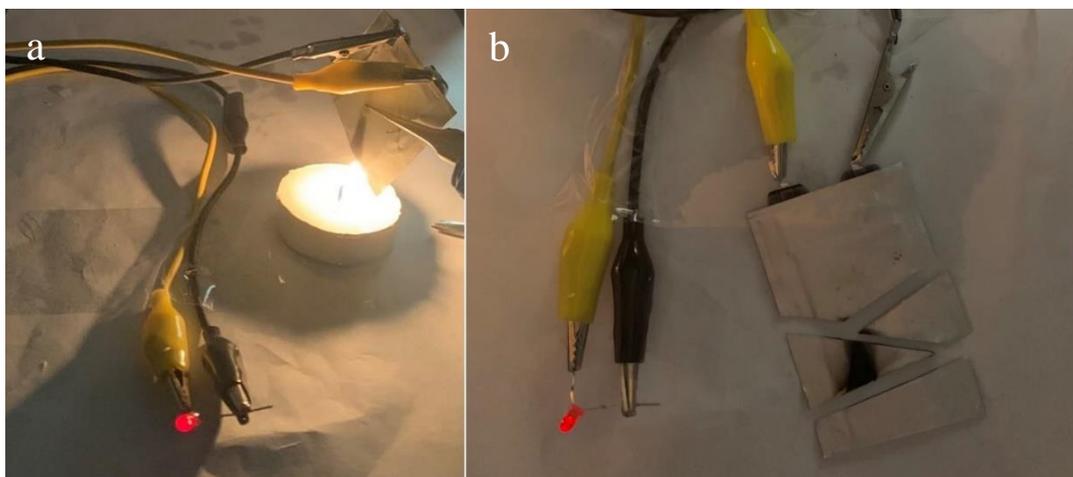

Fig. 5 (a) Flammability test of the GPE-based full battery. (b) Cutting test of the GPE-based full battery.

To verify the safety property of the full battery, we cut it into several pieces with scissors. During the cutting process, there is no leakage or fire caused by short circuit, and the battery still can light the LED after cutting (Fig. 5b). What' more, even after placing the fracture of the battery on the fire, the battery is still in good condition, without short circuit, fire, explosion or other dangerous situations, and the

LED works normally (Fig. 5a). The tests under these harsh conditions prove the superior safety of the GPE-based LFP/LTO full battery.

## Conclusion

In summary, high safety and cycle stability battery system using $Pyr_{13}FSI$/LiFSI/PVDF-HFP in LFP/LTO is developed. The low viscosity ionic liquid ($Pyr_{13}FSI$) and small size lithium salt (LiFSI) in the GPE reduce the interface impedance and greatly improve the ion mobility, which allows the battery to work at room temperature and deliver the specific capacity of 123 mAh/g (1C). Benefitting from the non-flammability and negligible vapour pressure of the ionic liquid, the battery can still work normally after being stored at 80 °C for ten hours. The wide electrochemical window of the GPE and the suitable working voltage of the LTO anode allow the battery to exhibit outstanding cycle stability, maintaining 80% of the initial capacity after 2000 cycles. Consequently, two major hazards faced by lithium batteries, flammable liquid electrolytes and the growth of lithium dendrites in anode, do not exist in our system and ultra-high safety is guaranteed.

1. Kim, T. H. *et al.* The current move of lithium ion batteries towards the next phase. *Adv. Energy Mater.* **2**, 860–872 (2012).

2. Guo, W. *et al.* Surface and Interface Modification of Electrode Materials for Lithium-Ion Batteries With Organic Liquid Electrolyte. *Front. Energy Res.* **8**, 1–20 (2020).

3. Ohsaki, T. *et al.* Overcharge reaction of lithium-ion batteries. *J. Power Sources* **146**, 97–100 (2005).

4. Spotnitz, R. & Franklin, J. Abuse behavior of high-power, lithium-ion cells. *J. Power Sources* **113**, 81–100 (2003).

5. Wang, Q., Sun, J., Yao, X. & Chen, C. Thermal Behavior of Lithiated Graphite with Electrolyte in Lithium-Ion Batteries. *J. Electrochem. Soc.* **153**, A329 (2006).

6. Finegan, D. P. *et al.* In-operando high-speed tomography of lithium-ion batteries during thermal runaway. *Nat. Commun.* **6**, 1–10 (2015).

7. Finegan, D. P. *et al.* In-operando high-speed tomography of lithium-ion batteries during thermal runaway. *Nat. Commun.* **6**, 1–10 (2015).

8. Armand, M. B. Polymer Electrolytes. *Annu. Rev. Mater. Sci.* **16**, 245–261 (1986).

9. Liang, S. *et al.* Gel polymer electrolytes for lithium ion batteries: Fabrication, characterization and performance. *Solid State Ionics* **318**, 2–18 (2018).

10. Li, W. *et al.* A PEO-based gel polymer electrolyte for lithium ion batteries. *RSC Adv.* **7**, 23494–23501 (2017).

11. Mathew, C. M., Kesavan, K. & Rajendran, S. Structural and Electrochemical Analysis of PMMA Based Gel Electrolyte Membranes. *Int. J. Electrochem.* **2015**, 1–7 (2015).

12. Sekhon, S. S., Arora, N. & Agnihotry, S. A. PAN-based gel electrolyte with lithium salts. *Solid State Ionics* **136–137**, 1201–1204 (2000).

13. Periasamy, P. *et al.* Studies on PVdF-based gel polymer electrolytes. *J. Power Sources* **88**,

269–273 (2000).

14. Jie, J. *et al.* High-performance PVDF-HFP based gel polymer electrolyte with a safe solvent in Li metal polymer battery. *J. Energy Chem.* **49**, 80–88 (2020).

15. Navarra, M. A. Ionic liquids as safe electrolyte components for Li-metal and Li-ion batteries. *MRS Bull.* **38**, 548–553 (2013).

16. Karuppasamy, K. *et al.* Ionic liquid-based electrolytes for energy storage devices: A brief review on their limits and applications. *Polymers* (2020) doi:10.3390/POLYM12040918.

17. Mousavi, M. P. S. *et al.* Ionic Liquids as Electrolytes for Electrochemical Double-Layer Capacitors: Structures that Optimize Specific Energy. *ACS Appl. Mater. Interfaces* **8**, 3396–3406 (2016).

18. Han, H. B. *et al.* Lithium bis(fluorosulfonyl)imide (LiFSI) as conducting salt for nonaqueous liquid electrolytes for lithium-ion batteries: Physicochemical and electrochemical properties. *J. Power Sources* (2011) doi:10.1016/j.jpowsour.2010.12.040.

19. Kerner, M., Plylahan, N., Scheers, J. & Johansson, P. Thermal stability and decomposition of lithium bis(fluorosulfonyl)imide (LiFSI) salts. *RSC Adv.* **6**, 23327–23334 (2016).

20. Guo, Z., Zhu, J., Feng, J. & Du, S. Direct in situ observation and explanation of lithium dendrite of commercial graphite electrodes. *RSC Adv.* **5**, 69514–69521 (2015).

21. Luo, J. *et al.* A proof-of-concept graphite anode with a lithium dendrite suppressing polymer coating. *J. Power Sources* **406**, 63–69 (2018).

22. Dimov, N., Kugino, S. & Yoshio, M. Carbon-coated silicon as anode material for lithium ion batteries: Advantages and limitations. *Electrochim. Acta* **48**, 1579–1587 (2003).

23. Park, M. G., Lee, D. H., Jung, H., Choi, J. H. & Park, C. M. Sn-Based Nanocomposite for Li-Ion Battery Anode with High Energy Density, Rate Capability, and Reversibility. *ACS Nano* **12**, 2955–2967 (2018).

24. Rao, X. *et al.* Polyacrylonitrile Hard Carbon as Anode of High Rate Capability for Lithium Ion Batteries. *Front. Energy Res.* **8**, 1–9 (2020).

25. Sun, X., Radovanovic, P. V. & Cui, B. Advances in spinel Li4Ti5O12 anode materials for lithium-ion batteries. *New J. Chem.* **39**, 38–63 (2015).

26. Huang, J. & Hollenkamp, A. F. Thermal behavior of ionic liquids containing the FSI anion and the Li + cation. *J. Phys. Chem. C* **114**, 21840–21847 (2010).

27. Zhou, Q., Henderson, W. A., Appetecchi, G. B., Montanino, M. & Passerini, S. Physical and electrochemical properties of N-alkyl-N-methylpyrrolidinium bis(fluorosulfonyl)imide ionic liquids: PY13FSI and PY 14FSI. *J. Phys. Chem. B* **112**, 13577–13580 (2008).